# Supersonic Ramjet Engine Inlet for Jovian Flight

Martin N. Karanikolov; Nedislav S. Veselinov; Dimitar M. Mladenov

# Abstract

This paper discusses the analysis performed on a supersonic ramjet engine inlet for flight in the atmosphere of Jupiter. Since the Jovian atmosphere lacks oxygen, the thrust will be generated by nuclear fission heating in the heat chamber. The first task to solve in the design in a ramjet engine is to design the supersonic inlet. The developed design methodology utilizes theoretical calculations and Computational Fluid Dynamics (CFD) simulations. The analytical model used to calculate the gas parameters in front of the heat chamber, and the CFD analysis, used to define the inlet geometry, are discussed. The results from the analytical model and CFD are compared and used for validation of the design approach. The calculated pressure losses and the mass flow allow the determination of important parameters required for the design of the aircraft, such as the reactor power, the thrust, the maximum mass, and the overall external dimensions.

# Table of Contents

# List of Figures

# List of Tables

# Abbreviations

| | | | |
|---|---|---|---|
| CFD | Computational Fluid Dynamics | NPRE | Nuclear-Powered Ramjet Engine |
| JSL | Jovian Sea Level | | |

# 1    Introduction

A Nuclear-Powered Ramjet Engine (NPRE) can be used to propel an aircraft (Flyer) in the Jovian atmosphere. Such engines have been studied and tested during the 50s and 60s during the Project Pluto in the United States [1]. During that time, the design means for developing ramjet engines and airbreathing aircraft were limited to theoretical calculations and ground and flight testing. Nowadays, simulation tools like Computational Fluid Dynamics (CFD) are available, allowing the design process to be vastly accelerated and conducted at a significantly lower cost.

In contrast to previous research on airbreathing engines, the analysis considers the unique environment on Jupiter and offers insight of how the engine performs in a low-temperature, hydrogen-rich atmosphere. A major challenge was to determine whether the CFD models (which are normally used for simulating flight in terrestrial conditions) are valid for the Jovian gas. To verify that, a series of CFD simulations for air and Jovian gas at different temperatures was conducted on a body with known aerodynamics, for which experimental data is readily available. The results are promising, showing that the resulting pressure distributions ($c_p$ distributions) are similar in all investigated cases. This is important since there is no way of recreating the conditions on Earth economically. Performing high-speed test flights and extensive wind tunnel testing with hydrogen gas is impossible, leaving simulation as the only viable means for the design of the engine and aircraft.

An overview of the design and previous work on the NPRE Flyer is presented in Section 2. The engine geometry and analytical calculations are discussed in Section 3. The results of the CFD validation study and the inlet design simulations are presented in Section 4.

# 2    Previous Work and Design Approach

In the initial study [2], the thrust requirement for a Flyer suitable for flight in the Jovian atmosphere was determined at different altitudes and heat chamber temperatures. The analysis relied on environmental data from the NASA Galileo probe [10] and envisaged an idealized NPRE. It was determined that a NPRE with 600 K heat chamber temperature will produce sufficient thrust to propel a 1000 kg Flyer at Mach 3 and altitude of around 60 km above JSL (the Jovian Sea Level (JSL) is defined as the isobaric surface of static pressure equal to the Earth's Mean Sea Level pressure or 1013.25 hPa).

Since the main goal was to determine the flight levels as a function of thrust, the model used an approximate thrust calculation approach, where compressibility effects and diffusion where completely omitted. In order to calculate the thrust and required reactor power accurately, a more exact calculation approach is needed.

The calculation method presented in Section 3 accounts for the supersonic gas compression throughout the engine. The analysis considers a one-dimensional supersonic shock system to calculate the thermodynamic gas parameters (pressure, temperature, etc.) explicitly. The thrust produced by the NPRE is determined as function of the selected altitude, engine cross-section and reactor power. The methodology is similar to the calculations described in [5, 8, 9], but is adapted and applied to Jupiter's extraterrestrial environment with its unique thermophysical properties, as reported by Galileo. The resulting engine geometry is optimized for the specific conditions on Jupiter.

Compared to turbojet engines, the NPRE features very simple design (Figure 1). During flight, the undisturbed flow (position 1) enters the engine through the supersonic inlet (2), where it is slowed down through a system of supersonic shocks. In this process, the static pressure, temperature and density of the gas increase across each successive shock. The gas, slowed down to subsonic speed, continues to diffuse in the subsonic diffusor (3) before it enters the heat chamber (h.c.), where it is heated up in the nuclear fission process to the design temperature. The hot, highly compressed gas enters the nozzle through the critical section (4cr), where it is accelerated to high supersonic speed, producing thrust (e). This simple design with few moving parts is advantageous for space exploration, since it offers good to weight to thrust ratio, does not require complicated lubrication, and features lower risk of mechanical failure.

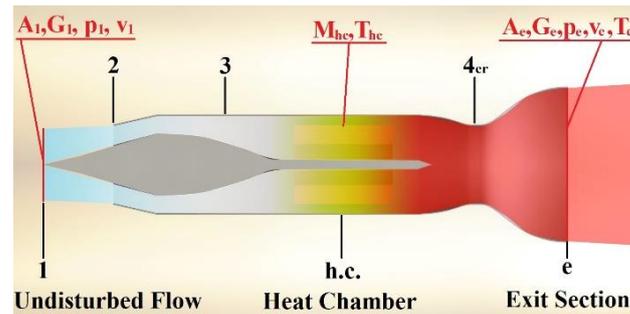

**Figure 1: Nuclear-Powered Ramjet Engine Layout**

To determine the required nuclear reactor power and the resulting thrust, the parameters of the gas (pressure, density, temperature and Mach number) entering the heat chamber need to be calculated. The supersonic diffusion process is associated with losses. There is unpreventable total pressure loss across the shock system, which depends on the inlet geometry and can be expressed by the pressure recovery coefficient [8]. At given engine cross-section, a lower-loss inlet with higher pressure recovery will produce more thrust.

The demonstrated inlet geometry is an intermediary result, subject to modifications as the Flyer design progresses and the performance requirements evolve. However, it represents a major milestone since the design demonstrates an optimal shock system with minimum additional drag and adequate performance for steady flight. The geometry can be scaled according to the evolving thrust requirements. Additionally, the results provide a cross-check of the methodology by demonstrating to what extent the CFD simulation results agree with the theoretical calculations.

During the development of the inlet, several geometries were investigated. The geometry presented in Section 4 offers high compression at high mass flow rate. The proposed inlet offers high thrust at minimal cross-section and weight.

The engine parameters provide the input necessary for designing the other craft components and allow the calculation of the outer dimensions and the payload mass.

# 3   Analytical Model

Among other factors, the thrust produced by ramjets depends on the angle of propagation of the shock waves produced by the "spike" of an axisymmetric (Figure 2, left) or the ramp of a rectangular (Figure 2, right) inlet. Ideally, the angles in a supersonic inlet should be optimized to produce a shock system that originates from the inlet outer leading edge (Figure 10, position 1) [3]. Such shock configuration will induce minimum additional drag. Early supersonic aircraft, like the F-104, were designed with non-adjustable inlets that operated at optimal condition only at the design (cruise) altitude and speed. At higher Mach numbers, excessive air was bypassed around the engine. Modern inlets exhibit axially adjustable central body or adjustable ramps (Concorde's Olympus engines being one example, Figure 2), which ensure the correct mass flow through the engine is maintained under most flight conditions.

The inlet losses / pressure recovery and thrust depend on the obliquity of the shocks. Supersonic inlets are designed with surfaces of incrementally increasing angles to produce a system of oblique (weak) shocks to gradually decrease the Mach number before the flow is brought down to subsonic speed through a normal shock [5, 7, 8]. Higher pressure recovery is achievable by increasing the number of oblique shocks (Figure 5).

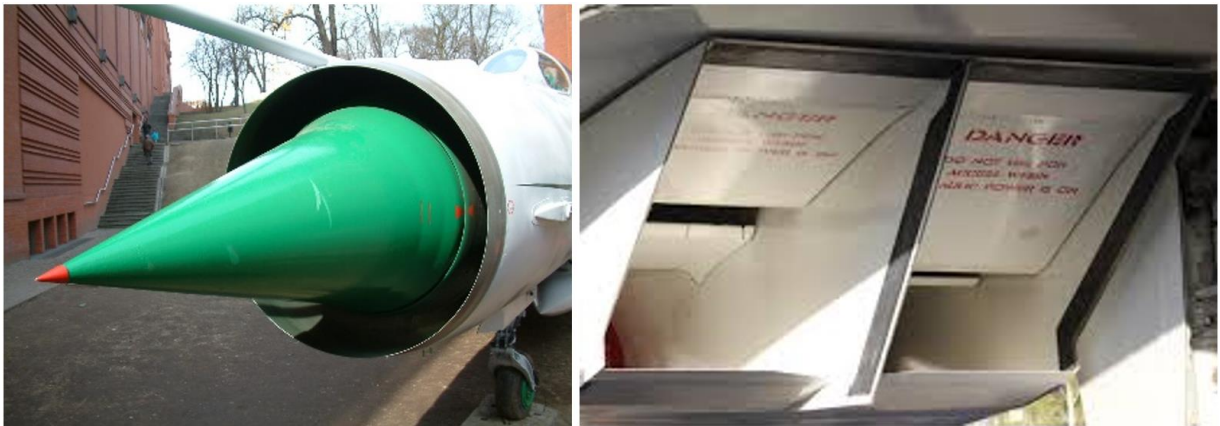

**Figure 2: Supersonic inlets of an early fighter aircraft (left) and the Concorde (right)**

This analysis considers one-dimensional flow through a fixed axisymmetric inlet with a system of three oblique shocks, followed by a normal shock and subsonic diffusor duct.

## 3.1   Flight Environment

The selected flight altitude for this analysis is 60 km above JSL. The thermophysical properties of the undisturbed flow per Galileo's data are as follows [2, 10]:

$$p_{u.f.} = 4374 \, Pa; \quad \rho_{u.f.} = 0.00991 \, kg/m^3; \quad T_{u.f.} = 122.6 \, K; \quad \gamma = 1.534$$

$$M_J = 2.309 \times 10^{-3} \, kg/mol; \quad R_J = 3600.8 \, \frac{J}{kg.K}$$

Where $\gamma$ is the heat capacity ratio, $M_J$ is the molar mass, and $R_J$ is the specific gas constant of the Jovian atmosphere.

The flight will take place in the stratospheric layer, where the probability for strong perturbations and high winds is lower (Figure 3). Additionally, the selected altitude is 30 km above the ammonia layer, which will improve the experimental capabilities of the mission by allowing unobstructed observation in the visible part of the spectrum.

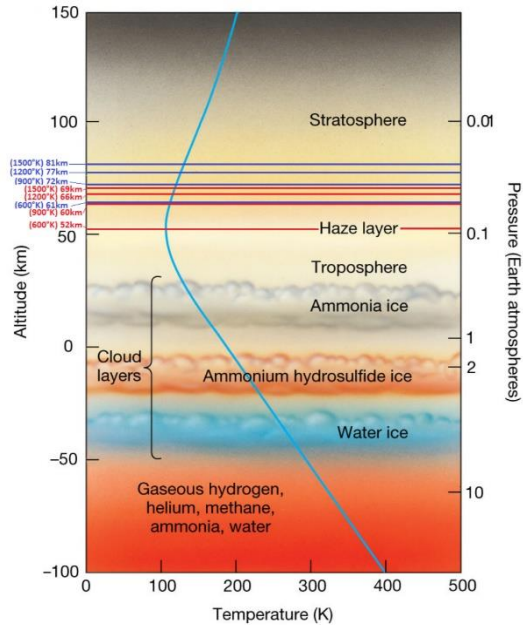

**Figure 3: Atmospheric composition of Jupiter (figure source https://www.uccs.edu/ / Pearson Education, Inc.)**

## 3.2 Calculations

The objective of the calculation is to determine the gas mass flow through the engine and the gas properties at the heat chamber inlet. The inlet portion of the engine is divided into sections 1-7. Section 1 is upstream of the supersonic inlet, where the flow is undisturbed (at ambient conditions), before being slowed down through the shock system 2-6 (Figure 4). At section 6, the flow enters the subsonic portion of the engine and, subsequently, the heat chamber (section 7). The calculated gas properties in front of the heat chamber allow to determine the required reactor power and engine thrust.

The relation between the upstream Mach number $M_i$, the angle of flow rotation $\omega_i$, and the shock obliquity angle $\beta_i$ through shock $i$ is as follows [5, 8]:

$$\frac{1}{M_i^2} = sin^2\beta_i - \frac{\gamma + 1}{2} \cdot \frac{\sin\beta_i . sin\omega_i}{\cos(\beta_i - \omega_i)}$$

The relation is used to calculate $\beta_i$ for given $\omega_i$ and $M_i$. Obviously, the flow rotation angles will correspond to the structural angles of the inlet (Figure 4).

The physical properties downstream of the shock $p_{i+1}, \rho_{i+1}, T_{i+1}, M_{i+1}$ and the pressure recovery $\sigma_i$ can be calculated from the following relations:

$$p_{i+1} = p_i \cdot \left(\frac{2.\gamma}{\gamma + 1} \cdot M_i^2 \cdot sin^2\beta_i - \frac{\gamma - 1}{\gamma + 1}\right)$$

$$\rho_{i+1} = \rho_i \cdot \left[\left(\frac{2}{\gamma + 1} \cdot \frac{1}{M_i^2 \cdot sin^2\beta_i} - \frac{\gamma - 1}{\gamma + 1}\right)^{-1}\right]$$

$$T_{i+1} = T_i \frac{p_{i+1}}{p_i} \cdot \frac{\rho_i}{\rho_{i+1}}$$

$$M_{i+1} = \frac{1}{\sin(\beta_i - \omega_i)} \cdot \sqrt{\frac{1 + \frac{\gamma - 1}{2} \cdot M_i^2 \cdot \sin^2\beta_i}{\gamma \cdot M_i^2 \cdot \sin^2\beta_i - \frac{\gamma - 1}{2}}}$$

$$\sigma_i = \frac{1}{\left(\frac{p_{i+1}}{p_i}\right)^{\frac{1}{\gamma-1}} \cdot \left(\frac{\rho_i}{\rho_{i+1}}\right)^{\frac{\gamma}{\gamma-1}}}$$

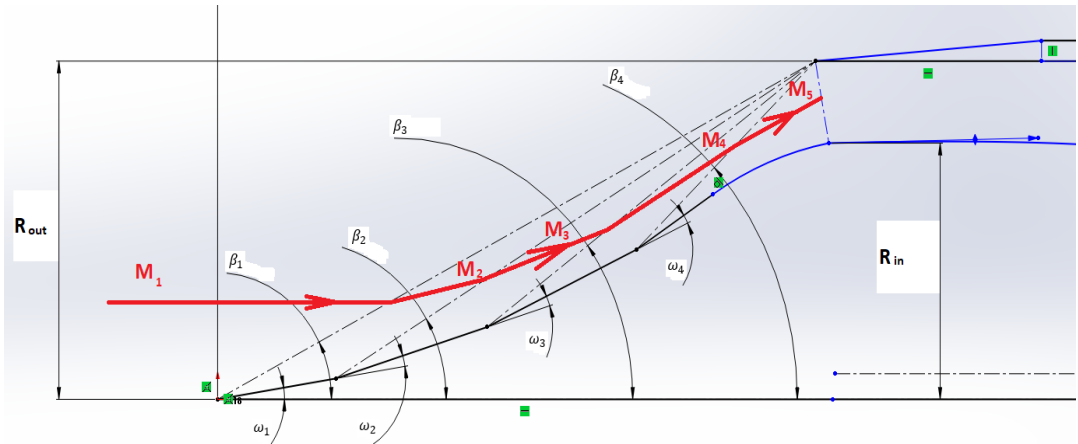

**Figure 4: Relation between shock wave and flow turning angles from CAD (earlier design)**

The pressure recovery coefficient is a measure for the total pressure loss through a shock. The stronger the shock, the lower the pressure recovery.

The shock system parameters of the inlet are given in Table 1. The fourth oblique shock was omitted ($\omega_4 = 0°$), which helps reduce the diameter of the engine considerably. This configuration offers good balance between inlet performance and the overall dimensions of the Flyer, as it considers the capabilities of the existing rocket carriers.

| $\omega_i$ | $\beta_i$ | $p_{i+1}$; $\rho_{i+1}$; $T_{i+1}$; $M_{i+1}$ | $\sigma_i$ |
|---|---|---|---|
| $\omega_1 = 9°$ | $\beta_1 = 27°$ | $p_2 = 8877 Pa$; $\rho_2 = 0.016\,kg/m^3$; $T_2 = 159\,K$; $M_2 = 2.47$ | $\sigma_1 = 0.97$ |
| $\omega_2 = 8°$ | $\beta_2 = 31°$ | $p_3 = 15296$; $\rho_3 = 0.022\,kg/m^3$; $T_3 = 193\,K$; $M_3 = 2.09$ | $\sigma_2 = 0.986$ |
| $\omega_3 = 8°$ | $\beta_3 = 36°$ | $p_4 = 24813 Pa$; $\rho_4 = 0.03\,kg/m^3$; $T_4 = 229\,K$; $M_4 = 1.76$ | $\sigma_3 = 0.99$ |
| $\omega_4 = 0°$ | $\beta_4 = -$ | $p_5 = p_4$; $\rho_5 = \rho_4$; $T_5 = T_4$; $M_5 = M_4$ | $\sigma_4 = 1$ |
| *Gas parameters downstream of the normal shock:* | | | |
| | | $p_6 = 87496 Pa$; $\rho_6 = 0.065\,kg/m^3$; $T_6 = 376\,K$; $M_6 = 0.64$ | $\sigma_5 = 0.844$ |

**Table 1: Analytically determined gas parameters at different sections of the engine inlet**

The total pressure recovery of the whole system is as follows:

$$\sigma_\Sigma = \sigma_1 \cdot \sigma_2 \cdot \sigma_3 \cdot \sigma_4 \cdot \sigma_5 = 0.80$$

The calculated total inlet pressure loss is 20%. This result agrees well with inlet losses of supersonic aircraft, built for flight on Earth. Figure 5 shows the relation between the achievable pressure losses and the flight Mach number for shock systems with different number of shocks [5]. "x" and "o" mark the result from this work and an airbreathing engine inlet respectively.

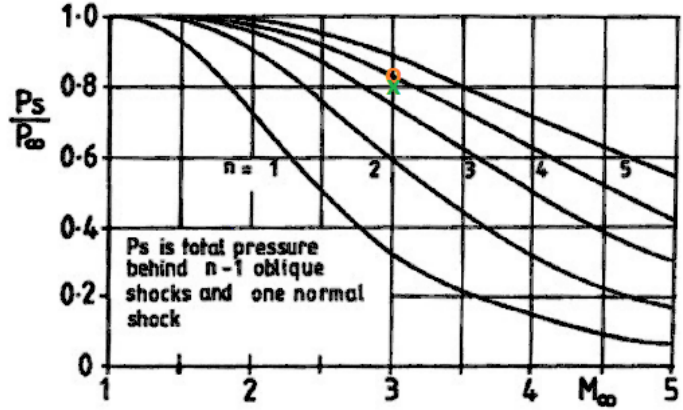

**Figure 5: Pressure recovery for different number of shocks and freestream Mach numbers**

The conclusion is that the pressure losses predicted by the analytical model are comparable to the losses stated by other sources [5].

The calculated gas parameters downstream of the normal shock (section 6 of the engine) are comparable to airbreathing ramjet engines and demonstrate the good performance of the inlet: relative to the undisturbed flow, the pressure is 20 times higher, the density has increased about 6.5 times, and the temperature has increased over 3 times. The flow is purely subsonic, which reduces the risk of shock waves developing in the heat chamber.

The following section 6-7 is the subsonic diffusor. The gas parameters at the diffusor inlet are as follows (Table 1):

$$p_6 = 87496 \; Pa; \quad \rho_6 = 0.065 \; kg/m^3; \quad T_6 = 376 \; K; \quad M_6 = 0.639$$

The gas diffuses further and reaches the following the parameters at the heat chamber inlet:

$$p_7 = 109\,576 \; Pa; \quad \rho_7 = 0.075 \; kg/m^3; \quad T_7 = 407 \; K; \quad M_7 = 0.309$$

For the investigated geometry with outer engine duct diameter of $d_{out} = 740 \; mm$ and maximal central body diameter of $d_{in} = 480 \; mm$ (dimensions of the narrowest engine section), the calculated mass flow is:

$$\dot{m} = 14.82 \; kg/s$$

The estimated reactor power required to heat up the flow to 600 K is about 38 MW. The engine will produce about 11 kN of thrust, which is sufficient for sustainable flight of a Flyer with about 1000 kg mass at the selected altitude. This estimate considers 20% thermal losses through the skin of the craft and the nozzle.

The method can be used to calculate the thrust and power at altitudes ranging from 24 km to 90 km above JSL and for different heat chamber temperatures.

# 4 CFD Study

During the initial design, different inlet geometries were studied by utilizing CFD. The objectives were (1) to define an inlet geometry, ensuring optimal operation at the selected altitude and flight speed, and (2) to verify the design methodology by comparing the flow parameters from the analytical model with the CFD results.

In contrast to other work, where only the internal (ducted) flow is simulated [5], the Jovian Flyer inlet is modeled in a large domain that includes the external volume (Figure 6). This leads to more accurate simulation of the shock system at given flight conditions, since it allows for full development of the flow structures inside and around the front of the engine.

## 4.1 CFD Model Overview

The selected physical models are typical for high-speed simulations. The simulation utilizes coupled RANS solver and the *k-ω* (SST) turbulence model. The atmosphere is modelled as non-reacting two-component gas mixture with 87% Hydrogen and 13% Helium composition by mass fraction.

The simulation model used for the geometry optimization is two-dimensional. The engine inlet is placed in a large volume with boundaries at sufficient distance from the inlet surfaces and includes a long downstream duct extension. The model is meshed with a polyhedral mesh with ~40000 cells. The mesh includes prism layers along the surfaces and local mesh refinements where the shock system is formed.

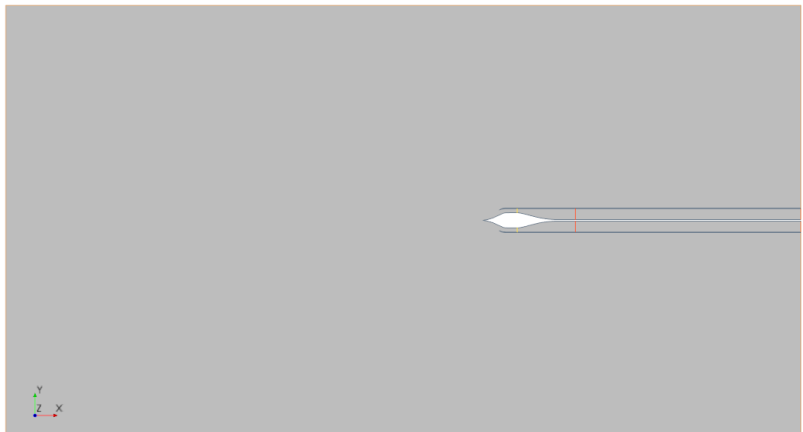

**Figure 6: CFD simulation domain and simulated body**

The boundary conditions correspond to the Galileo data for 60 km and are identical to the conditions used in the analytical model: the undisturbed flow is at 122.6 K temperature and 4374 Pa static pressure. The outside domain walls are free flow boundaries with gas flowing at M = 3.0 parallel to the engine axis. The thermophysical properties of the two-component gas are heat capacity ratio $\gamma = 1.534$, molar mass $M_J = 2.309 \times 10^{-3} \ kg/mol$ and specific gas constant of $R_J = 3600.8 \frac{J}{kg.K}$. The inlet is simulated without the heat chamber and other engine components. Any downstream components will introduce resistance or blockage, affecting the flow. To simulate the presence of the heat chamber, a porous barrier (membrane) with a pressure drop coefficient of 2 is defined downstream of the inlet. The gas flows through the barrier before leaving the domain at sufficient downstream distance.

## 4.2 CFD Model Validation Against Experimental Data

The selected CFD models are widely used for simulating highspeed aerodynamics in Earth's atmosphere. To verify whether they are also valid for the Jovian gas, a series of CFD simulations for air and Hydrogen-Helium gas at different temperatures was conducted on a body with known aerodynamics, for which wind tunnel data is readily available.

In his work [11] Petrov investigates the pressure distributions around bodies with simple shapes. The work is a compilation of experimental results on $c_p$ distributions around test articles in supersonic flow. A three-dimensional CFD model of a test article was developed and simulated by using the physical models, intended for the inlet simulation. The model was run for air at room temperature and for Jovian gas, and the results were compared against the experimental (wind tunnel) data.

The geometry and $c_p$ distributions at different Mach numbers of Body 7.16 from Petrov's work are shown in Figure 7 on the left. The information allowed the construction a 3D model of the body and to run CFD simulations to generate the $c_p$ distributions, as shown in Figure 7, right. The locations along the body surface where $c_p$ is probed, are seen in red.

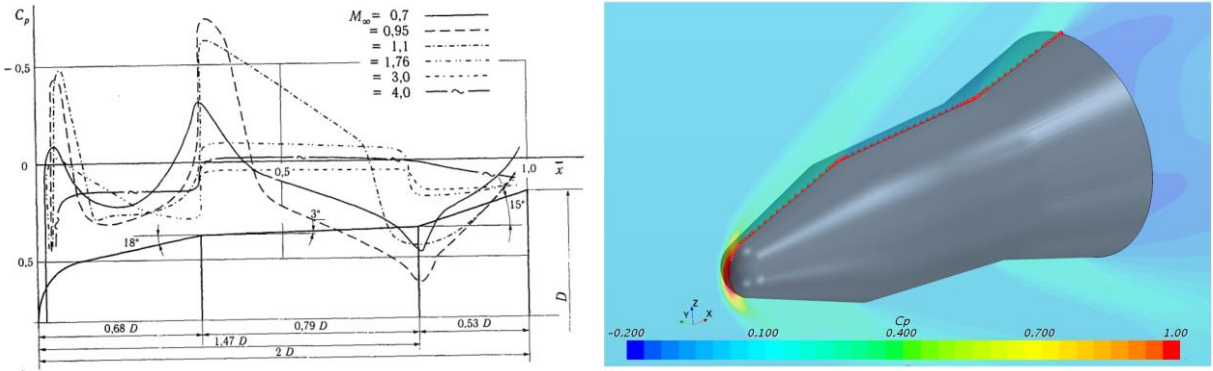

**Figure 7:** $c_p$ distribution around body 7.16 from [11] at zero angle of attack

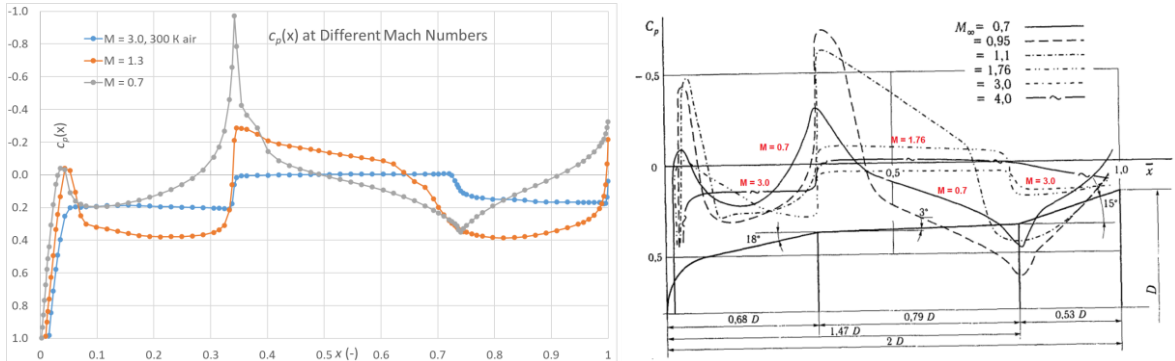

**Figure 8:** $c_p$ distribution comparison from CFD (left) and [11] (right) at different Mach numbers

The simulated $c_p$ distributions along the axis are shown in Figure 8, left. Comparing the M = 3.0 line (blue) and M = 0.7 line (grey) with Petrov's data (–·– line and continuous line, respectively) shows that the $c_p$ distributions are reproduced accurately at both supersonic and subsonic speeds. At M = 0.7, a small deviation is observed around the front convex transition, where the CFD model predicts stronger suction due to flow acceleration. This can be attributed to a difference between Petrov's measurement locations and where the CFD solution is probed. However, the $c_p$ distributions at the supersonic speeds of interest (M = 3.0) along the flat surfaces are both qualitatively and quantitatively a very good match.

Petrov's data is for air at room temperature. It is essential to evaluate how the $c_p$ behavior changes for Jovian gas. Figure 9 shows comparison between $c_p$ for air at room temperature (blue), air at 122 K (orange), and for H2-He gas at 122 K (grey), all simulated with the designated CFD model.

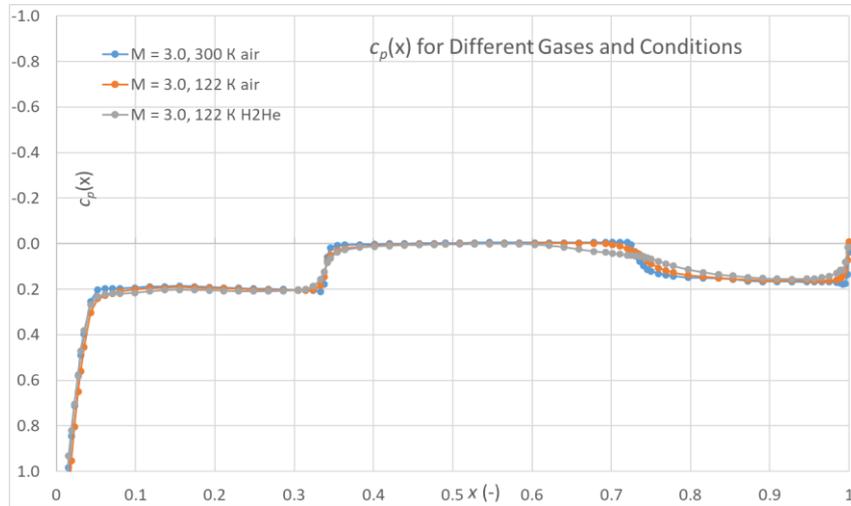

**Figure 9: $c_p$ distribution on body surface at M = 3.0 for air and two-component Jovian gas at different temperatures**

It is apparent that despite the difference in Re numbers, the predicted $c_p$ distributions are nearly identical over most of the surface. Small deviations are observed around the concave transition at the back (coordinate $0.6 < x < 0.8$), which can be attributed to the higher viscosity of the fluid at low temperatures.

The result of this comparison is of high significance, since it shows that the flow pressure behavior at given Mach number is similar in the different gas environments on Earth and Jupiter. This is a very encouraging result, given how difficult and costly would be to recreate these conditions to allow for testing during the development of the Flyer.

The results show that the gas behavior can be predicted by means of CFD with high accuracy, which grants higher confidence in the development methodology.

## 4.3 Flyer Inlet CFD Results

### 4.3.1 Flow Distributions

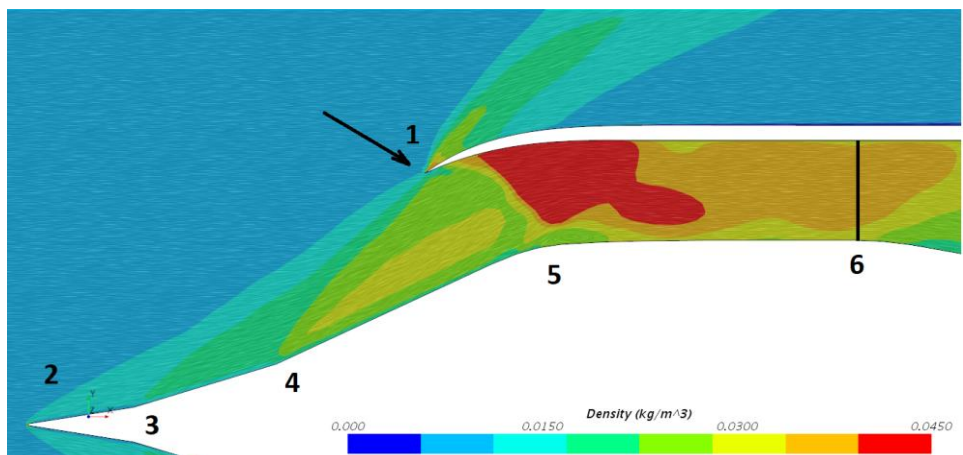

**Figure 10: The simulated density distribution clearly shows the shock structure**

The inlet simulation shows that at the design altitude and speed, the shock wave structure originates from a common point on the inlet annular leading edge, as intended. Figure 10 illustrates the density distribution around the inlet. It is evident that the individual oblique shocks are formed between the annular leading edge (position 1) and the points where the flow changes direction (2, 3, 4). Such shock wave pattern suggests that the inlet is functioning optimally, with the anticipated wave drag being minimal [3].

The flow diffusion is clearly apparent: downstream of every shock, the density increases incrementally. The curved outer leading edge and central body geometries are selected carefully to form slightly converging cross-section in the upstream section of the duct. This causes slight blockage which leads to a formation of a strong normal shock at position 5. The flow passes the narrowest section of the engine (throat) and enters the subsonic diffusor at 6.

The Mach number distribution along the axis of the engine is shown in Figure 11. The Mach number gradually decreases from 3.0 in the free flow to 0.85 immediately downstream of the normal shock. However, further downstream the flow accelerates to 1.4 and a supersonic jet is formed along the downstream section of the central body. After analysis of the results, it was concluded that the high-resistance porous membrane cannot substitute the effect of an explicitly modelled heat chamber and other downstream components.

Regardless of the inaccurate subsonic diffusion simulation, the objective of determining the inlet geometry and optimizing the supersonic shock system and upstream section of the engine was achieved.

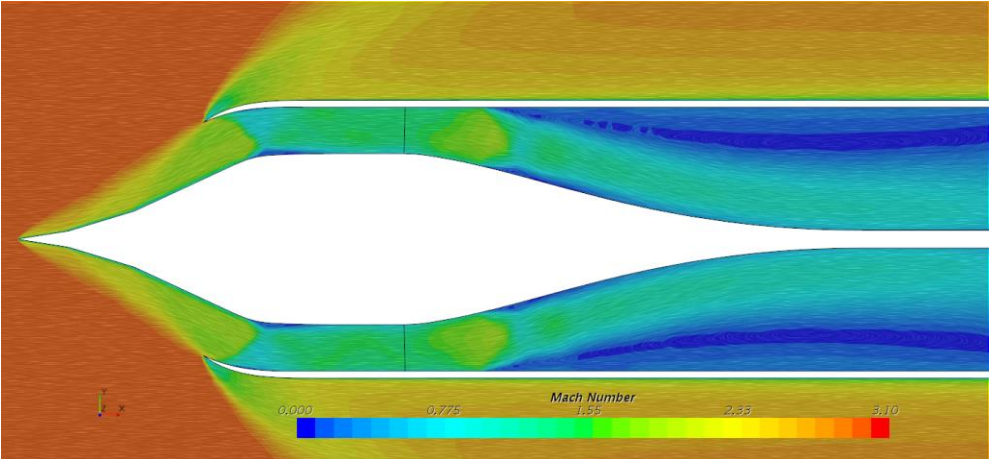

**Figure 11: Mach number distribution throughout and around the inlet**

### 4.3.2  Gas Properties – Comparison Between Analysis and CFD

The second objective was to compare the flow properties from CFD with the calculations from the analytical model. To compare the results directly, the flow parameters need to be evaluated at identical sections of the engine.

The analytical model is one-dimensional, and the properties are calculated at different engine sections. However, the CFD model is two-dimensional, and the value of the probed parameter will depend on where the CFD solution is probed. For ducted flow, there is the choice between calculating the averaged parameters at different cross-sections or to probe at a single location at half-height of the duct, for example. However, the outside flow around the inlet front is not ducted and there are no solid outside boundaries, so averaging is meaningless. For this reason, the solution needs to be probed at discrete locations.

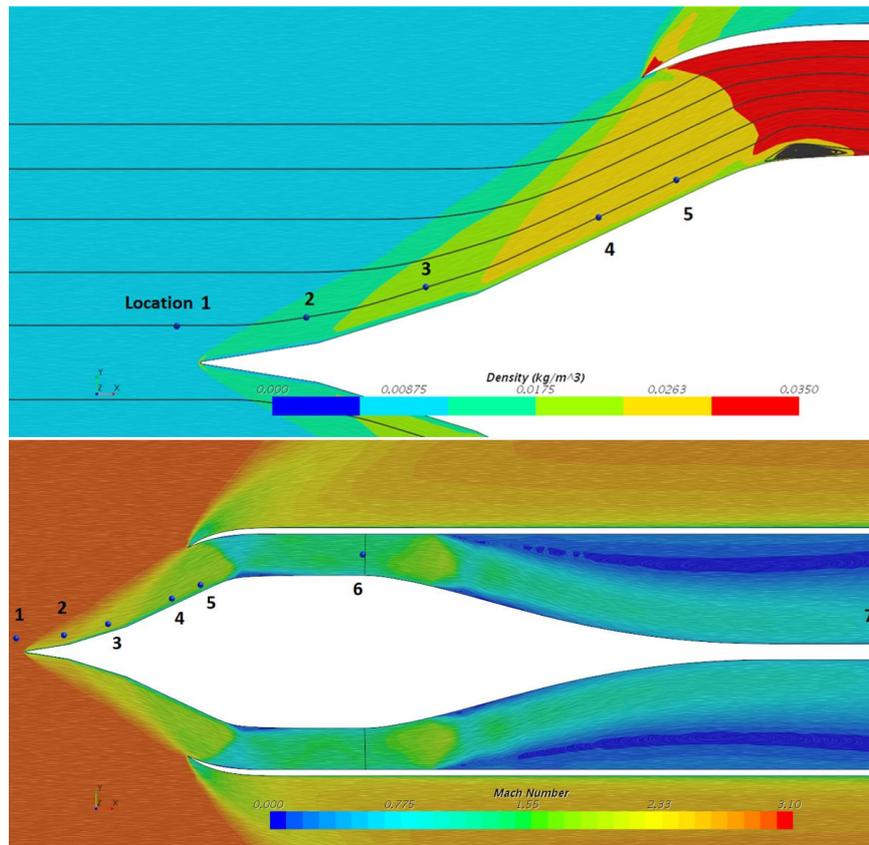

**Figure 12: Solution probing locations for gas parameter comparison between the analytical model and CFD**

After analyzing the CFD data, it was determined that the flow parameters between the shocks do not change considerably streamwise. Significant gradients are observed only in direction, normal to the model surface. For this reason, the solution was probed along a streamline at halfway between two neighboring shocks (Figure 12, above). In the ducted portion, the probing locations are chosen at half-height of the duct (Figure 12, below). These locations are away from the boundary layer and are preferred since the boundary layer effects need to be excluded.

The results of the comparison are presented in Figure 13 to Figure 16. All gas parameters, as predicted by the analytical calculations and CFD, are in good agreement in the supersonic section of the inlet 1-5. The observed disagreement in the subsonic diffusion section is caused by the limitations of the CFD model, which does not include the downstream components required to accurately simulate the flow resistance. The lack of heat chamber allows the flow to re-accelerate, while the analytical calculation assumes that the flow is subject to increased blockage and subsonic diffusion.

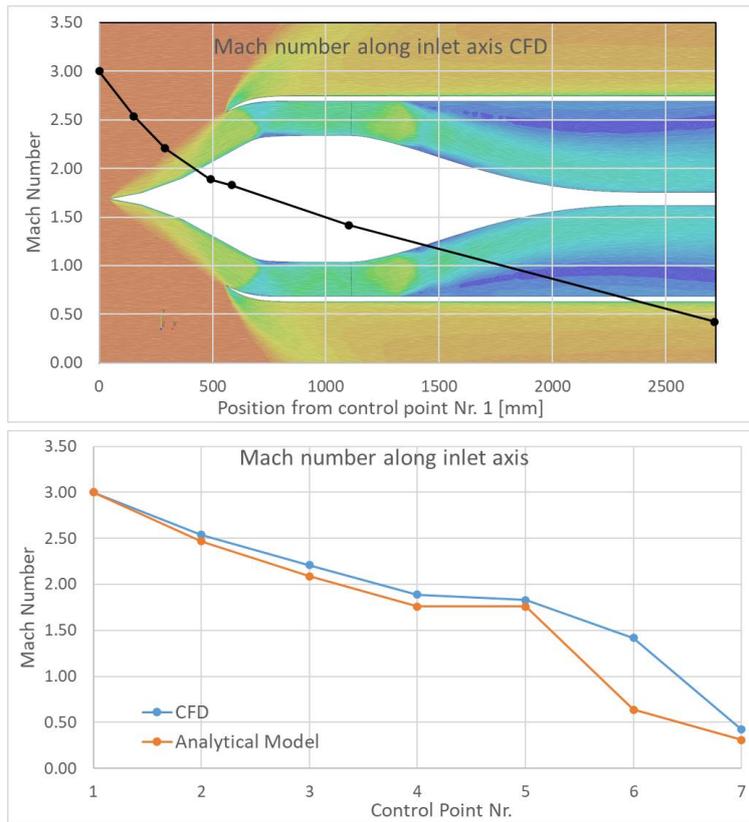

**Figure 13: Mach number along the engine axis from CFD and analytical model**

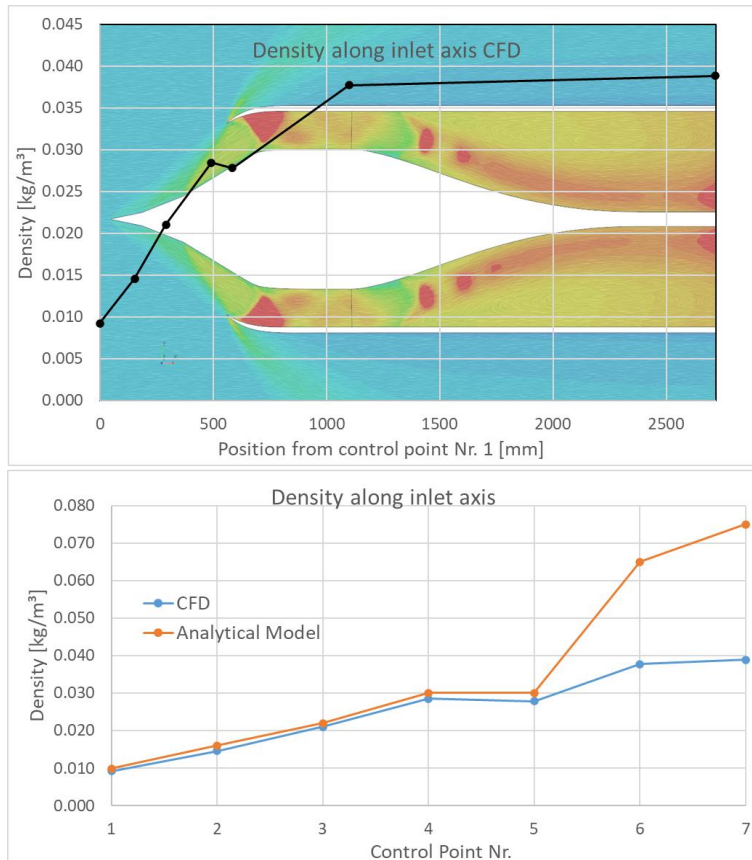

**Figure 14: Density along the engine axis from CFD and analytical model**

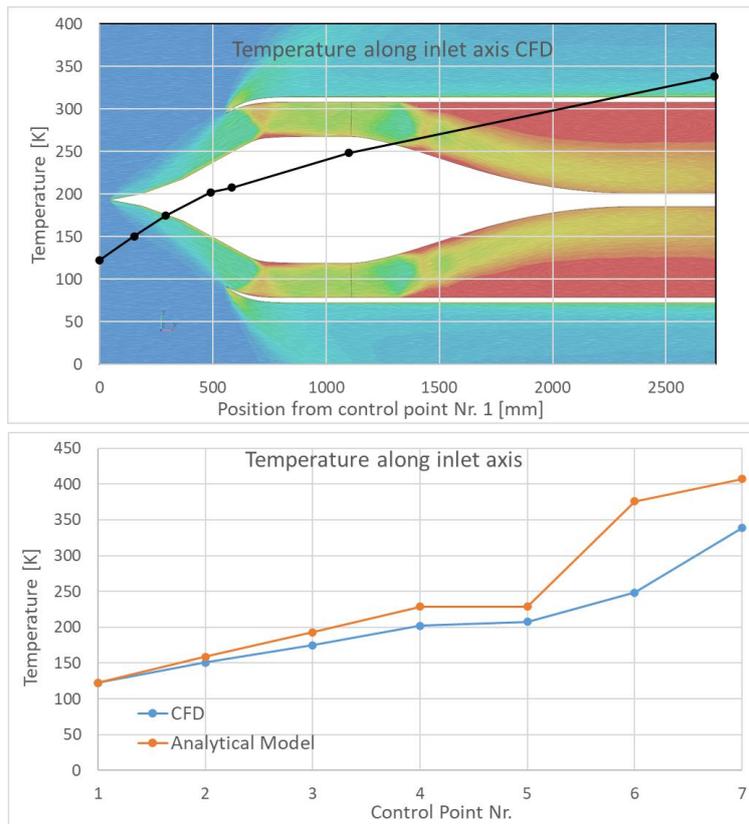

**Figure 15: Temperature along the engine axis from CFD and analytical model**

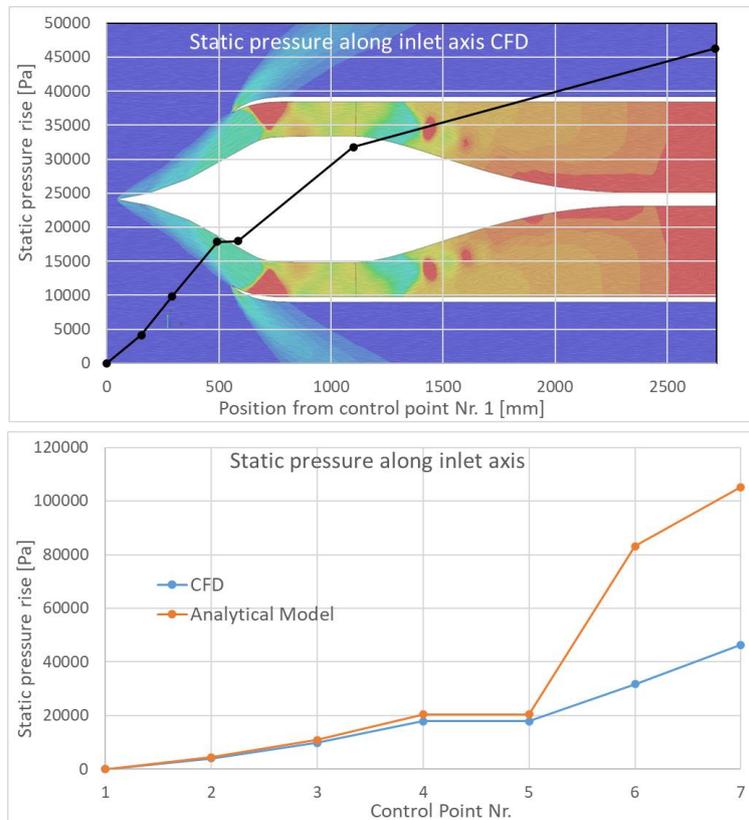

**Figure 16: St. pressure along the engine axis from CFD and analytical model. Increase over 4374 Pa**

### 4.3.3 Mass Flow Calculation

The final objective was to evaluate the mass flow. The mass flow will be identical in the different sections, as the mass conservation law needs to be respected. A good indicator whether mass is conserved along the engine axis is to compare the Density × Mach number product at different cross-section. This product represents the "mass flow per unit area" and allows a direct comparison between the two models, including the unducted front inlet section.

The Density × Mach number product in different sections is shown in Figure 17. It is apparent that the CFD results are in good agreement with the theoretical calculations.

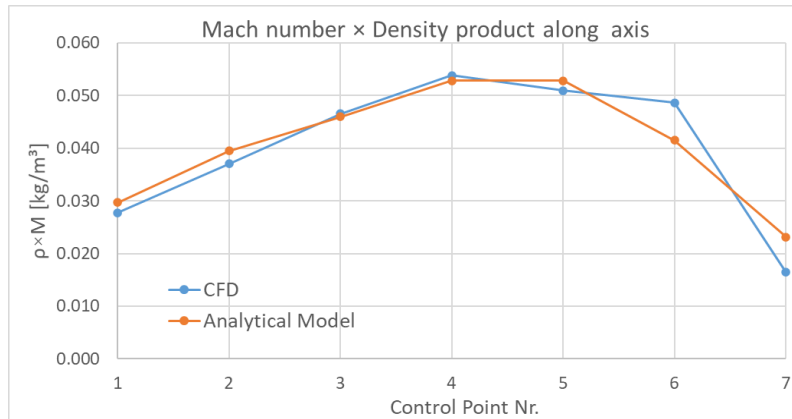

**Figure 17: Density × Mach number product along the engine axis**

Figure 18 shows the convergence of the mass flow during the simulation in different sections (the throat, at the heat chamber inlet and at the outlet of the model):

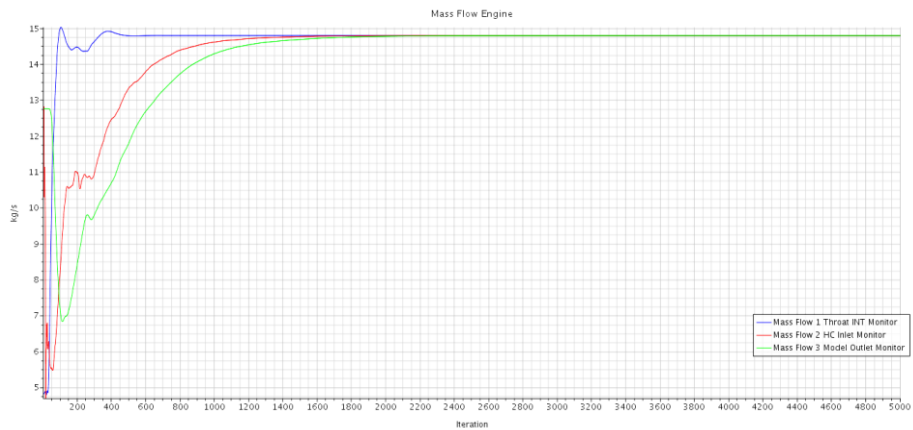

**Figure 18: Mass flow convergence during 5000 calculation iterations in different engine sections**

The good convergence of the solution to the exact same value demonstrates that mass conservation is satisfied. The reported mass flow is identical up to the sixth decimal position. The CFD model yields:

$$\dot{m} = 14.79 \frac{kg}{s}$$

An overview of the gas parameters at different sections is shown in Table 2. The analytically calculated mass flow agrees with the result from the CFD simulation within 0.3%. Compared to the analytically predicted, the losses from CFD are higher (25.8% vs. 20.1%). This is because the 2D simulation also reproduces the external flow, while the analysis assumes purely one-dimensional flow, not considering the patterns outside the duct. An additional shock forms at the outer leading edge of the engine as shown on Figure 11, which contributes to the additional losses.

|  | Mach Number | | Static Pressure [Pa] | | Density [kg/s] | | Temperature [K] | |
|---|---|---|---|---|---|---|---|---|
| Position | Analysis | CFD | Analysis | CFD | Analysis | CFD | Analysis | CFD |
| 1 | 3.00 | 3.00 | 4374 | 4374 | 0.0099 | 0.0092 | 122.6 | 122.6 |
| 2 | 2.47 | 2.54 | 8877 | 8476 | 0.0160 | 0.0146 | 159.0 | 150.5 |
| 3 | 2.09 | 2.21 | 15296 | 14207 | 0.0220 | 0.0211 | 193.0 | 174.8 |
| 4 | 1.76 | 1.89 | 24813 | 22216 | 0.0300 | 0.0285 | 229.0 | 202.2 |
| 5 | 1.76 | 1.83 | 24813 | 22314 | 0.0300 | 0.0278 | 229.0 | 207.7 |
| 6 | 0.64 | 1.42 | 87496 | 36172 | 0.0650 | 0.0377 | 376.0 | 248.6 |
| 7 | 0.31 | 0.42 | 109576 | 50696 | 0.0750 | 0.0388 | 407.0 | 338.3 |

|  | Analysis | CFD |
|---|---|---|
| Pressure recovery σ | 0.80 | 0.74 |
| Inlet losses, % | 20.1% | 25.8% |
| Mass flow, kg/s | 14.82 | 14.79 |

**Table 2: Comparison of the gas parameters, pressure recovery and resulting mass flow between the analytical model and CFD**

# 5   Conclusion and Future Work

The proposed analytical model allows the determination of the engine size and geometry, as well as the calculation of the required reactor power and resulting thrust.

The analytical model predicts pressure recovery comparable to the performance for airbreathing engines. The calculations are in good agreement with the supersonic inlet calculations performed by others.

The conducted theoretical and CFD analyses have achieved three major goals: a non-airbreathing engine inlet for extraterrestrial flight was developed with the geometry demonstrated achieving optimum performance; the CFD and analytical model were compared to each other, showing good agreement and that the methods can be used for the development of the supersonic inlet; the employed CFD physical models were verified by comparing the simulation data with experimental data from wind tunnels.

The validation study led to two highly encouraging conclusions: air at room temperature and Jovian gas at low temperatures interact in similar way with the aircraft surfaces; and, the simulated pressure distributions from CFD are very similar to the experimentally measured $c_p$ distributions, meaning CFD simulations can be used for the development of extraterrestrial aircraft for flight in non-aerobic atmospheres with high confidence.

The CFD model has proven to have limitations. In order to simulate the subsonic section accurately, the heat chamber needs to be modelled explicitly. Simulation of downstream parts of the engine will be subject of future work.

The results from the study provide the boundary conditions needed for the development of the further components and the external configuration of the Flyer. The determination of the thrust and pressure distributions around the aircraft enable the detailed design of the Flyer. Furthermore, it is possible to determine more accurately the flight altitudes and identify mission scenarios, the scientific payload and communications systems, and to define the carrier rocket requirements.

The analysis shows that the NPRE technology can be used for flight in other planets' atmospheres. The developed methodology provides a tool for the design of aircraft capable of flight in extraterrestrial environment and can be applied to any celestial body with thick atmosphere. The approach can be used to determine possible flight altitudes for flight on Venus, Saturn, Titan, Uranus and Neptune, and on known exoplanets.

# A     Works Cited

[1] Frank E. Rom, Analysis of a Nuclear-Powered Ramjet Missile, Research memorandum, Lewis Flight Propulsion Laboratory, Cleveland, Ohio,

(https://digital.library.unt.edu/ark:/67531/metadc60302/m2/1/high_res_d/19930088171.pdf)

[2] Veselinov. N.S., Karanikolov M.N. Shishkin V.V. Mladenov D.M.; *Flight in the Jovian Stratosphere – Engine Concept and Flight Altitude Determination*, Journal of Spacecraft and Rockets (Accepted for publication)

[3] Бондарюк М. М., Ильяшенко С. М., *Прямоточные воздушно-реактивные двигатели*, Москва, 1958

[4] Münzberg H. G., *"Flugantriebe"*, Springer-Verlag, 1972

[5] H. Ran and D. Mavris, "*Preliminary Design of a 2D Supersonic Inlet to Maximize Total Pressure Recovery*", AIAA 5th Aviation, Technology, Integration, and Operations Conference (ATIO), 26 - 28 September 2005, Arlington, Virginia

[6] Bussard R.W. and DeLauer R.D., *"Nuclear Rocket Propulsion - McGraw-Hill Series in Missile and Space Technology"*, 1958

[7] Sutton G.P. and Biblarz O., "*Rocket Propulsion Elements*", Ninth Edition, Wiley, 2017

[8] W.H.T. Loh, *"Jet, Rocket, Nuclear, Ion and Electric Propulsion: Theory and Design - Applied Physics and Engineering"*, An International Series, Volume 7, Springer-Verlag 1968

[9] Зуев Ю.В., Лепешинский И.А. - Приближенный газодинамический расчет сверхзвукового прямоточного воздушно-реактивного двигателя – 2009

[10] NASA PDS: The Planetary Atmospheres Node, Galileo Probe Data Set Archive (2003). URL: https://pds-atmospheres.nmsu.edu/PDS/data/gp_0001/

[11] Петров К.П. – Аэродинамика тел простейших форм – ISBN: 5-88688-014-3 – Физматлит 1998